\documentclass[conference]{IEEEtran}
\IEEEoverridecommandlockouts

\ifCLASSOPTIONcompsoc
\else
\fi

\ifCLASSINFOpdf
\else
\fi

\usepackage{authblk}

\usepackage{amsmath, amsthm}
\usepackage{algorithmic}
\usepackage[ruled,linesnumbered,lined,boxed,commentsnumbered]{algorithm2e}
\usepackage{graphicx}
\usepackage{fancyhdr}
\usepackage{cite}

\usepackage[hidelinks]{hyperref}
\usepackage{algorithmic}
\usepackage{xcolor}

\usepackage{titlesec}
\titlespacing\section{0pt}{3pt}{1pt}
\titlespacing\subsection{0pt}{2.5pt}{1pt}

\newtheorem{lemma}{Lemma}

\def\BibTeX{{\rm B\kern-.05em{\sc i\kern-.025em b}\kern-.08em
    T\kern-.1667em\lower.7ex\hbox{E}\kern-.125emX}}

\begin{document}

\title{Resource Allocation of Federated Learning Assisted Mobile Augmented Reality System in the Metaverse}

% \author{\IEEEauthorblockN{Xinyu Zhou, Yang Li}
% \IEEEauthorblockA{\textit{School of Computer Science and Engineering \& ERI@N} \\
% \textit{Nanyang Technological University}\\
% Singapore \\
% xinyu003@e.ntu.edu.sg\\
% yang048@e.ntu.edu.sg}
% \and
% \IEEEauthorblockN{Jun Zhao}
% \IEEEauthorblockA{\textit{School of Computer Science and Engineering} \\
% \textit{Nanyang Technological University}\\
% Singapore \\
% junzhao@ntu.edu.sg}
% }

%

\author[1]{Xinyu~Zhou}
\author[1]{Yang~Li}
\author[2]{Jun Zhao}
\affil[1]{\small  School of Computer Science \& Engineering and ERI@N, Nanyang Technological University, Singapore}
\affil[2]{\normalsize School of Computer Science \& Engineering, Nanyang Technological University, Singapore} 
\affil[ ]{\normalsize xinyu003@e.ntu.edu.sg, yang048@e.ntu.edu.sg, junzhao@ntu.edu.sg} 

% \markboth{Journal of \LaTeX\ Class Files,~Vol.~14, No.~8, August~2015}%
% {Shell \MakeLowercase{\textit{et al.}}: Bare Demo of IEEEtran.cls for Computer Society Journals}

\IEEEtitleabstractindextext{%
\begin{abstract}
Metaverse has become a buzzword recently.
Mobile augmented reality (MAR) is a promising approach to providing users with an immersive experience in the Metaverse.
However, due to limitations of bandwidth, latency and computational resources, MAR cannot be applied on a large scale in the Metaverse yet. 
Moreover, federated learning, with its privacy-preserving characteristics, has emerged as a prospective distributed learning framework in the future Metaverse world.
In this paper, we propose a federated learning assisted MAR system via non-orthogonal multiple access for the Metaverse.
Additionally, to optimize a weighted sum of energy, latency and model accuracy, a resource allocation algorithm is devised by setting appropriate transmission power, CPU frequency and video frame resolution for each user.
Experimental results demonstrate that our proposed algorithm achieves an overall good performance compared to a random algorithm and greedy algorithm.
\end{abstract}

\begin{IEEEkeywords}
Resource allocation, federated learning, augmented reality, Metaverse, NOMA.
\end{IEEEkeywords}}

% \pagestyle{empty} \thispagestyle{empty}
\maketitle

% \thispagestyle{fancy}
% \pagestyle{fancy}
% \lhead{This paper appears in the Proceedings of IEEE International Conference on Communications (ICC) 2023.\\ Please feel free to contact us for questions or remarks.}

%  \thispagestyle{fancy}
% \pagestyle{fancy}
% \lhead{}

\IEEEdisplaynontitleabstractindextext

\IEEEpeerreviewmaketitle

\section{Introduction}\label{sec:introduction}
Metaverse has become a buzzword in recent years. It seeks to create a society integrated virtual/augmented reality and allows millions of people to communicate online with a virtual avatar.
Augmented Reality (AR) technology, which has been expected to one of the most significant component of the Metaverse, is an enhanced version of the real physical world that is achieved through the use of digital visual elements, sound, or other sensory stimuli and delivered via technology. Mobile Augmented Reality (MAR) is to implement AR technology on mobile devices and allow users to experience services through AR devices (e.g. smart glasses, headsets, controllers, etc.).

% With rapid growth of mobile devices and the volume of mobile data \cite{andrews2014will, chang2017cluster,li20145g,chen2016time,he2018mcast}, non-orthogonal multiple access (NOMA) has been proposed as an effective solution in 5G system. In contrast to the conventional orthogonal multiple access (OMA), it introduces an extra power domain and allows different users to be multiplexed on the same channel to maximize the throughput of the system \cite{benjebbour2013concept}. Specifically, it uses superposition coding at the transmitter, and applies successive interference cancellation (SIC) at the receivers to differentiate signals from multiple users in the power domain \cite{he2019joint}.

\textbf{Motivation}.
According to Moore's law, the storage capacity and computing power of mobile devices will be further improved in the future, making it possible to implement machine learning models on mobile devices \cite{dionisio20133d}. However, limited by the small amount of personal data, it is difficult to train a high-performing MAR model on a single device. Federated learning (FL) \cite{mcmahan2017communication}, presented in 2017, allows models from diverse participants to train a global model with better performance while protecting user privacy. With FL, each device only needs to do local training with its own data and upload the model parameters to the server without sharing any information with other devices. The server will aggregate the parameters from different participants, form a better-performing global model and send it back for further training. Therefore, it is worth investigating how to use FL to enhance the MAR-based Metaverse experience.

Besides, non-orthogonal multiple access (NOMA) was proposed as an effective solution in 5G system \cite{li20145g,chen2016time,he2018mcast}. In contrast to the conventional orthogonal multiple access (OMA), it introduces an extra power domain. It allows different users to be multiplexed on the same channel to maximize the throughput \cite{benjebbour2013concept}. Specifically, it uses superposition coding at the transmitter and applies successive interference cancellation (SIC) at the receivers to differentiate signals from multiple users in the power domain. Hence, with the help of NOMA, we devise an FL-assisted MAR system in the Metaverse.

%\vspace{-10pt}
\begin{figure}[!t]
\centering
\includegraphics[width=9cm]{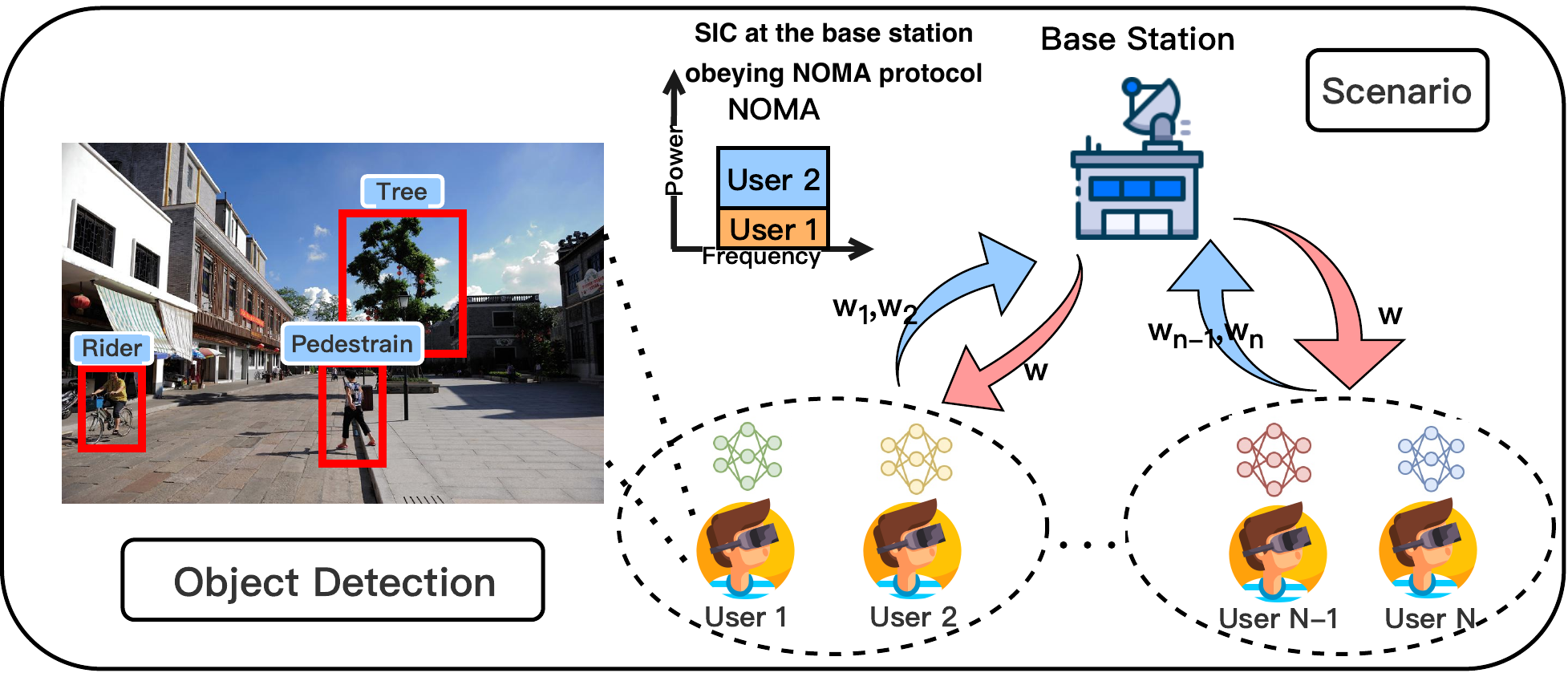}\vspace{-5pt}
\caption{Mobile
augmented reality (MAR) with NOMA for the Metaverse.}
\vspace{-15pt}
\label{fig:MAR_network}
\end{figure}

\textbf{Challenges}.
It still faces several challenges to deploy FL to MAR applications in the Metaverse: (1) Each device is often given limited bandwidth, so the upper bound of transmission rate is affected according to Shannon's formula, which causes long delay during the communication period, further influences the convergence of global model. (2) The massive energy consumption caused by the training of local model is a challenge for the power supply of mobile devices. (3) Video frame resolution is also a critical factor in the accuracy of the MAR model. Training with higher resolution could improve the performance but also means higher computational resource consumption for participants. Therefore, how to adjust video frame resolutions to balance model performance and total consumption is also a problem to ponder. 
% Although the use of NOMA could increase bandwidth utilization and improve communication quality to some extent, there is still much room for optimizing the settings of parameters (e.g. CPU frequency, transmission power, etc.) to further reduce the total consumption while ensuring model accuracy. 

\textbf{Related work}.
%% resource allocation and MAR
To improve the quality of experience in MAR applications, some studies tackled the resource allocation problem to utilize limited resources and achieved good fairness performance \cite{SONG2022, 7906521,9319727, chen2020joint, si2022resource,10.1007/978-3-030-01168-0_40}. \cite{si2022resource} designed an MAR-based model for the Metaverse and proposed a transmission resource allocation algorithm to allocate transmission power and resolution for each device to achieve the best utility. 
%% NOMA and MEC
NOMA had been deployed massively in multi-user mobile edge computing (MEC) networks to improve the communication resource, and quality of service \cite{li2019joint,ye2020enhance,gupta2022noma}. For instance, \cite{li2019joint} considered dynamic user pairing NOMA-based offloading and established an energy consumption minimization framework by joint optimizing pairing. 
% \cite{ye2020enhance} proposes a new hybrid offloading scheme in a NOMA-based MEC network that can operate in three different
% modes.
% \cite{gupta2022noma} designs a algorithm that utilizes the bottleneck matching algorithm, convex optimization, and the block coordinate
% descent scheme to obtain a locally-optimal solution in NOMA based MEC network. 
There is few work incorporating FL into MAR, and we only found one paper \cite{chen2020federated}. 
An FL-based mobile edge computing paradigm was proposed in \cite{chen2020federated} to solve object recognition and classification problem without considering optimizing the resource allocation in the framework.

\textbf{Novelty}. 
The previously mentioned studies \cite{li2019joint,ye2020enhance,gupta2022noma} were only about NOMA and MEC systems without integrating FL and MAR. \cite{chen2020federated} was about FL and AR without considering resource allocation. Additionally, although \cite{SONG2022, 7906521,9319727, chen2020joint, si2022resource,10.1007/978-3-030-01168-0_40} were related to resource allocation and MAR, they did not apply FL to their systems.
In this paper, we consider a basic FL-assisted MAR system via NOMA, and design an algorithm to jointly optimize time, energy and model accuracy simultaneously. 
Besides, we take into account the different requirements in diverse situations. For example, we prefer to optimize energy consumption as much as possible when the mobile device is low on power.
So, the objective function includes a weighted combination of total energy consumption, completion time and model accuracy and allows the weights to be adjusted to achieve different optimization results.

\textbf{Contributions}. The contributions of our paper are as follows:
\begin{itemize}
    \item To our best knowledge, we are the first to introduce FL assisted NOMA system in MAR to enhance the model performance, which could improve the experience of users in the Metaverse.
    \item An optimization algorithm is proposed to optimize time, energy, and accuracy jointly. The weights of them could be adjusted freely according to practical situations.
    \item Detailed comparative experiments, convergence analysis and time complexity, are provided to show the robustness and effectiveness of our method.
\end{itemize}

\section{System Model}
We consider an MAR network with $N$ cellular-connected  MAR devices. All MAR devices transmit data to a BS through the uplink NOMA transmission protocol. Each device trains its own object detection model locally and collaboratively trains a global model through FL, as shown in Fig. \ref{fig:MAR_network}. In this paper, bold symbols represent vectors. If a symbol $x$ is a solution to a problem, then $x^*$ means the optimal solution.

\textbf{Uplink NOMA}. In an uplink-NOMA system, through channel and power assignment, signals of devices are multiplexed on each channel. Assume there are $N$ users and $K$ channels. The higher the number of devices multiplexed on each channel, the higher the hardware complexity and latency. Thus, following \cite{zhang2017downlink, he2019joint} and considering practicality, we suppose there are $2$ users multiplexed on the $k$-th subchannel. 

Suppose a user $n$ belongs to the subchannel $k$, and it is the $i$-th ($i=1,2$) user in the subchannel $k$. Define $g_{k,i}$ as the channel gain between the base station and the $i$-th device on channel $k$. Without loss of generality, assume $g_{k,i-1} < g_{k,i}$ on channel $k$ and the information of the user with better channel gain will be decoded first by the base station.

\textbf{User pairing}. Similar to \cite{ye2020enhance}, we consider using the following user pairing schemes: (a) random selection: randomly selecting two users to form a group, (b) nearest-user-pairing: pairing two nearest users, then the next nearest users, and so forth, (c) nearest-farthest-pairing: the nearest user and the farthest user are formed into a group, then pairing the second nearest and the second farthest user, and so forth. In the resource allocation algorithm (Algorithm \ref{algo:re_al}) stated in section \ref{subsec:re_al}, three user pairing schemes are used, and the best one will be selected as the final result.

\textbf{Federated learning}. Assume there are $D_n$ (i.e., $D_{k,i}$) samples on each device $n$ (i.e., the $i$-th device on channel $k$). As shown in Fig. \ref{fig:MAR_network}, each mobile device runs its own model locally and collaboratively solves the problem $\min_{\boldsymbol{\omega}}F(\boldsymbol{\omega})=\sum_{n=1}^N \frac{D_n}{\sum_{n=1}^N D_n}l_n(\boldsymbol{\omega})$, where $l_n(\cdot)$ denotes the loss function per sample and $\boldsymbol{\omega}$ is the global model.
After each device finishes its local training for a certain number of iterations, it will upload the model parameter to the base station. Next, the base station will calculate the weighted average model parameter $\frac{D_n\boldsymbol{\omega}_n}{\sum_{n=1}^N D_n}$ and send it back to each device. 
Such uploading and broadcasting process is called one \textit{global communication round}.

\vspace{-2pt}
\subsection{Energy and Time Consumption}
We only study the process between two global communication rounds. We define the \textit{total energy consumption} as $\mathcal{E}$, and it includes wireless \textit{transmission energy} and \textit{local computation energy}. The \textit{total time consumption} is defined as $\mathcal{T}$. It consists of \textit{transmission time} and \textit{local computation time}. Following \cite{dinh2021federated, yang2021energy}, the energy consumed at the base station is not considered.

\textbf{Transmission energy}. In light of the fact that the base station's output power is significantly greater than the uplink transmission power of a mobile device, the downlink time is ignored in this work. Hence, according to the Shannon formula, the data transmission rate of the $i$-th user on channel $k$ is
\begin{align}
   \hspace{-2pt} r_{k,i} \!=\! B_k\!\log_2(\!1 \!+\! \frac{p_{k,i}g_{k,i}}{B_kN_{k}+{\sum_{j=0}^{i-1}}p_{k,{j }}g_{k,{j}}}\!),
\end{align}
and we define $p_{k,0}=0$. We denote the total bandwidth is $B$, and $B_k = \frac{B}{K}$. $p_{k,i}$ refers to the transmission power. $g_{k,i}$ is the channel between the base station and the $i$-th device on channel $k$. $N_k$ is the noise power spectral density of Gaussian noise. Suppose the transmission data size of each device is $d_{k,i}$, so the transmission time of the $i$-th user on channel $k$ is 
\begin{align} \label{equa:trans_t}
    T_{k,i}^{trans} = d_{k,i}/r_{k,i}.
\end{align}
Therefore, the corresponding transmission energy is
\begin{align} \label{equa:trans_e}
    E_{k,i}^{trans} = p_{k,i}T_{k,i}^{trans}.
\end{align}

\textbf{Local computation energy}. We incorporate You Only Look Once (YOLO) algorithm \cite{redmon2016you} to handle the object detection tasks on each device. Note that the main structure of YOLO is convolutional neural network (CNN). Assume on the $i$-th device on channel $k$, the video frame resolution used for training is $s_{k,i}\times s_{k,i}$ pixels. Thus, due to the influence of the frame resolution used for training on computing resources \cite{krizhevsky2012imagenet} and motivated by \cite{mao2016dynamic}, the local computation energy is defined as
\begin{align} \label{equa:cmp_e}
    E_{k,i}^{cmp} = \kappa\eta\xi s_{k,i}^2c_{k,i}D_{k,i}f_{k,i}^2,
\end{align}
where $\kappa$ represents the effective switched capacitance, $\eta$ is the number of local iterations, $s_{k,i}^2$ is the pixels of the frame resolution, $D_{k,i}$ is the number of samples, $f_{k,i}$ is the CPU frequency and $c_{k,i}$ is the number of CPU cycles per \textit{standard sample}. 
We define the \textit{standard sample} as a video frame with a resolution of $s_0 \times s_0$ pixels and $\xi$ = $\frac{1}{s_0^2}$.
This means if a video frame with $s_{k,i}^2$ pixels and $s_{k,i}=s_0$, the local computation energy will be $\kappa\eta c_{k,i}D_{k,i}f_{k,i}^2$, which is the same as the definition in \cite{mao2016dynamic}. 

Hence, the total energy consumption $\mathcal{E}$ is 
\begin{align}
    \mathcal{E} = \sum_{k=1}^K\sum_{i=1}^2(E_{k,i}^{trans} + E_{k,i}^{cmp}).
\end{align}

\textbf{Transmission time}. This is given in Eq. (\ref{equa:trans_t}).

\textbf{Computation time}. The local computation time {\color{black} (i.e., the local training time)} of the $i$-th device on channel $k$ in one global iteration is
\begin{align} \label{equa:cmp_t}
    T_{k,i}^{cmp} = \eta\frac{\xi s_{k,i}^2c_{k,i}D_{k,i}}{f_{k,i}}.
\end{align}
Thus, the total completion time is
\begin{align}
\mathcal{T} \!=\! \max \{T_{k,i}^{trans} \!+\! T_{k,i}^{cmp}\},~ i=1,2, ~k \in [1,K].
\end{align}

\vspace{-2pt}
\subsection{Accuracy Analysis}
Denote $\mathcal{A}$ be the training accuracy of the whole federated learning process and define it as a function of $s_{k,i}$, which is $\mathcal{A}(s_{1,1}, s_{1,2},\cdots , s_{K,1},s_{K,2}) = \sum_{k=1}^K\sum_{i=1}^2 A_{k,i}$. 

The accuracy model from \cite{liu2018edge} is used in this work. Based on YOLO algorithm, \cite{liu2018edge} constructs the accuracy function regarding different video frame resolutions. Thus, the accuracy function of the $i$-th user on channel $k$ is defined as
\begin{align} \label{accuracy_func}
A_{k,i} = 1-1.578e^{-6.5\times 10^{-3}s_{k,i}}.
\end{align}

\vspace{-2pt}
\section{Joint Optimization of Energy, Time and Accuracy with Fixed Users}
\vspace{-2pt}
In this section, problem formulation, problem decomposition and solutions to the optimization problem will be illustrated, respectively. 

\vspace{-5pt}
\subsection{Problem Formulation}
A joint optimization of energy $\mathcal{E}$, time $\mathcal{T}$ and accuracy $\mathcal{A}$ problem is formulated in this section.
The optimization problem is as follows:
\begin{subequations}\label{problem:origin}
\begin{align} 
        \hspace{-10pt} \min_{s_{k, i}, ~f_{k, i}, ~p_{k, i}} & \alpha \mathcal{E} +\beta \mathcal{T} - \gamma \mathcal{A},\tag{\ref{problem:origin}} \\
 \text{subject to,}~
    & p^{min} \le p_{k,i} \le p^{max}, ~k\in [1, K], ~i=1,2,\label{constra:p_range}\\
    & \hspace{-10pt} f^{min} \le f_{k,i} \le f^{max}, ~k\in [1, K], ~i=1,2,\label{constra:f_range}\\
    & \hspace{-10pt} s_{k,i} \in \{s_1, s_2, s_3\},\label{constra:resolution_range}
\end{align}
\end{subequations}
where $s_{k,i}$, $f_{k,i}$ and $p_{k,i}$ are three optimization variables. $\alpha$, $\beta$ and $\gamma$ are three weight parameters, and $\alpha+\beta=1$, $\alpha, \beta \in [0,1]$, and $\gamma \ge 0$.
Constraints (\ref{constra:p_range}) and (\ref{constra:f_range}) limit the ranges of the transmission power and CPU frequency of each device. Constraint (\ref{constra:resolution_range}) sets three choices for the video frame resolution and $s_1 < s_2 < s_3$.

Due to the max function of $\mathcal{T}$, problem (\ref{problem:origin}) is \mbox{non-convex} and difficult to be decomposed. To avoid this difficulty, an auxiliary variable $T$ is introduced, and the problem becomes:
\begin{subequations}\label{problem:origin2}
\begin{align} 
    \min_{s_{k, i}, f_{k, i}, p_{k, i},T} & \alpha (\sum_{k=1}^K \sum_{i=1}^2 E_{k,i}^{cmp} + E_{k,i}^{trans}) +\beta T - \gamma \mathcal{A},\tag{\ref{problem:origin}} \\
    \text{subject to,} ~
    &(\ref{constra:p_range}),~(\ref{constra:f_range}),~(\ref{constra:resolution_range}) \notag\\
    &\hspace{-20pt} T_{k,i}^{trans}\!+\!T_{k,i}^{cmp}\!\le\!T,~i\in \{1,2\},~k=1,\cdots\!,\!K, \label{constra:T}
\end{align}
\end{subequations}
where constraint (\ref{constra:T}) is considered an upper bound of the total time consumption.

\subsection{Problem Decomposition}
Due to the original optimization problem (\ref{problem:origin}) being \mbox{non-convex} and quite complex, we split it into two subproblems (\textit{SP1} and \textit{SP2}) to make it easier to solve. Since $p_{k,i}$ only appears in $E_{k,i}^{trans}$, two subproblems are constructed, one with the optimization variable $f_{k,i}$ and $s_{k,i}$, and the other with $p_{k,i}$.

Subproblems 1 and 2 write as follows:

\begin{align}\label{Subproblem1}
 \textit{\textbf{SP1}:~~}\min_{s_{k,i}, f_{k,i},T} &\alpha (\sum_{k=1}^K \!\sum_{i=1}^2 E_{k,i}^{cmp}) \!+\! \beta T \!-\! \gamma \mathcal{A},\\
    \text{subject to},~& (\ref{constra:f_range}),~(\ref{constra:resolution_range}),~(\ref{constra:T}).\notag
\end{align}
\begin{align} \label{problem:subp2}
\textit{\textbf{SP2}:~~}\min_{p_{k,i}} &~\alpha\sum_{k=1}^K\sum_{i=1}^2 \frac{p_{k,i}d_{k,i}}{r_{k,i}},\\
    \text{subject to},~
    & (\ref{constra:p_range}),~(\ref{constra:T}).\notag
\end{align}

\subsection{Solution to SP1}
Since the video frame resolution $s_{k,i}$ is discrete, we relax it into a continuous variable $\hat{s}_{k,i}$ to make \textit{SP1} easier to tackle. Next, we handle $A_{k,i}(\hat{s}_{k,i})$.
Because our research mainly focuses on the application of federated learning in NOMA, so we introduce a simple but effective linear approach to construct the function $A_{k,i}$ (\ref{accuracy_func}). By using two points $(s_1,A_{k,i}(s_1))$ and $(s_3,A_{k,i}(s_3))$,we approximate $A_{k,i}(\hat{s_{k,i}})$ as the following linear function $A_{k,i}(\hat{s}_{k,i})$:

\begin{align}
    \hat{A}_{k,i}(\hat{s}_{k,i})=\hat{k}_{k,i}(\hat{s}_{k,i}-s_1)+A_{k,i}(s_1),
\end{align}
where $\hat{k}_{k,i} = \frac{A_{k,i}(s_3)
-A_{k,i}(s_1)}{s_3-s_1}$.

Then, the \textit{SP1} becomes:
\begin{subequations}\label{More_Simplified:Subproblem1}
\begin{align}
    \min_{s_{k,i}, f_{k,i},T} &\alpha \sum_{k=1}^K\! \sum_{i=1}^2 E_{k,i}^{cmp} \!+\! \beta T \!-\!
    % \notag\\
    \gamma \sum_{k=1}^K \sum_{i=1}^2\hat{A}_{i,k}(\hat{s}_{k,i})
    \tag{\ref{More_Simplified:Subproblem1}} \\
    \text{subject to},
    & ~(\ref{constra:f_range}),~(\ref{constra:T}),
    \notag\\
    & s_1 \leq \hat{s}_{k,i} \leq \hat{s}_3,
    ~\forall k \in [1,K],~i=1,2.\label{constraint:s}
\end{align}
\end{subequations}
It is easy to verify that the function of \textit{SP1} is convex, and the constraints are also. Karush-Kuhn-Tucker (KKT) approach works well to get the optimal solutions for this optimization problem. 
% Treat the \textit{SP1} by KKT approach and we can introduce the Lagrange function:

% \begin{align}
%     &L_1(f_{k,i},\hat{s}_{k,i},T,\lambda) \!=\! \alpha 
%     \sum_{k=1}^K\sum_{i=1}^2 \kappa \eta \xi \hat{s}_{k,i}^2c_{k,i}D_{k,i}f_{k,i}^2 \!+\! \beta T \notag\\
%     &\!-\! \gamma \sum_{k=1}^K \!\sum_{i=1}^2\hat{A}_{k,i}
%     (\hat{s}_{k,i}) \!+\! \sum_{k=1}^K \!\sum_{i=1}^2\lambda_{k,i}[(T_{k,i}^{cmp}\!+\!T_{k,i}^{up})\!-\!T], 
% \end{align}

% where $\boldsymbol{\lambda}:=[\lambda_{1,1},\lambda_{1,2},\lambda_{2,1},...,\lambda_{k,1},\lambda_{k,2}]$ is the multiplier associated with the inequality constraint.
After applying KKT conditions,
% $(\frac{\partial{L_1}}{\partial{f_{k,i}}}=0, \frac{\partial{L_1}}{\partial{\hat{s}_{k,i}}}=0, \frac{\partial{L_1}}{\partial{T}}=0, \lambda_{k,i}[(\eta\frac{\xi\hat{s}_{k,i}^2c_{k,i}D_{k,i}}{f_{k,i}}+T_{k,i}^{up})-T]=0)$,
we get 

% \begin{align}
%     \frac{\partial{L_1}}{\partial{f_{k,i}}} & \!=\!  2\alpha \kappa_{k,i}\eta \xi \hat{s}_{k,i}^2c_{k,i}D_{k,i}f_{k,i} \!-\!
%     \lambda_{k,i}\eta\frac{\xi \hat{s}_{k,i}^2c_{k,i}D_{k,i}}{f_{k,i}^2}\!=\!0, 
%     \\
%      \frac{\partial{L_1}}{\partial{\hat{s}_{k,i}}} & =2\alpha
%     \kappa_{k,i}\eta \xi \hat{s}_{k,i}c_{k,i}D_{k,i}f_{k,i}^2+
%     \lambda_{k,i}\eta\frac{2\xi\hat{s}_{k,i}
%     c_{k,i}D_{k,i}}{f_{k,i}}
%     \notag\\
%     &-\gamma\hat{k}_{k,i}=0, \label{KKT:s_n}
%     \\
%     \frac{\partial{L_1}}{\partial{T}} &=\beta - \sum_{k=1}^K
%     \sum_{i=1}^2\lambda_{k,i}=0, \label{KKT:T}
%     \\
%     \lambda_{k,i}&[(\eta\frac{\xi\hat{s}_{k,i}^2c_{k,i}D_{k,i}} 
%     {f_{k,i}}+T_{k,i}^{up})-T]=0, 
% \end{align}

% from which we derive

\begin{align}
    \textstyle{f_{k,i}^*} &\!=\! \textstyle{\sqrt[3]{\frac{\lambda_{k,i}}{2\alpha
    \kappa}}},~  
    \textstyle{\hat{s}_{k,i}^*}\!=\!\textstyle{\frac{\gamma\hat{k}_{k,i}}{2\eta\xi c_{k,i}
    D_{k,i}(\alpha\kappa f_{k,i}^2+\frac{\lambda_{k,i}}{f_{k,i}})}},
    \label{relation_f_s_lamda}\\
    \textstyle{\beta} &= \textstyle{\sum_{k=1}^K\sum_{i=1}^2\lambda_{k,i}}, \label{beta_lambda}
\end{align}
where $\lambda_{k,i}$ is the Lagrange multiplier associated with the inequality constraint (\ref{constra:T}).

Through Eq. (\ref{relation_f_s_lamda}), we can use $\lambda_{k,i}$ to represent $f_{k,i}$ and $\hat{s}_{k,i}^2$. 

Then, we can get the dual problem as follows:

\begin{subequations}\label{subproblem1_dual}
\begin{align}
    \max_{\lambda_{k,i}}\sum_{k=1}^K\sum_{i=1}^2 
    &-\frac{\gamma^2\hat{k}_{k,i}^2}{4h(2^{-\frac{2}{3}}+2^{\frac{1}{3}})}\lambda_{k,i}^{-\frac{2}{3}}+T_{k,i}^{up}\lambda_{k,i} + \gamma\hat{k}_{k,i}s_1 \notag\\
    &- \gamma A_{k,i}(s_1)
    \tag{\ref{subproblem1_dual}} 
    \\
    \text{subject to},&~(\ref{beta_lambda}),~\lambda_{k,i} \geq 0, \notag
\end{align}
\end{subequations}
where $h=\eta\xi c_{k,i}D_{k,i}(\alpha \kappa )^{\frac{1}{3}}$. Obviously, this dual problem is a simple convex optimization problem. In this paper, we use CVX \cite{grant2014cvx} to solve it and get the optimal $\lambda^*=[\lambda_{1,1}^*,...,\lambda_{k,2}^*]$. Then we can leverage the $\lambda^*$ to calculate the optimal $f^*$ and $\hat{s}^*$ through Eq. (\ref{relation_f_s_lamda}). With the constraint of $f^{min} \le f_{k,i} \le f^{max}$, we can get $f_{k,i}^*$ in the below:

\begin{align}
    f_{k,i}^*=\min(f^{max},\max(f_{k,i}^*,f^{min})).
\end{align}

Since the frame resolution $s_{k,i}$ is discrete, we adopt the following formula to map $\hat{s}_{k,i}$ to $s_{k,i}$:

\begin{align}
    s_{k,i}^* = 
    \begin{cases}
    s_3, ~\text{if} ~\hat{s_{k,i}} > \frac{s_2+s_3}{2}\\
    s_2, ~\text{if} ~\frac{s_1+s_2}{2} \leq \hat{s_{k,i}}
    \leq \frac{s_2+s_3}{2}\\
    s_1, ~\text{if} ~\hat{s_{k,i}} < \frac{s_1+s_2}{2}
    \end{cases}
\end{align}
\textit{SP1} is solved. Next, \textit{SP2} will be explained.

\vspace{-5pt}
\subsection{Solution to SP2}

The objective function (\ref{problem:subp2}) of \textit{SP2} is \mbox{non-convex}, since it is easy to verify that its Hessian matrix is not positive semidefinite. On channel $k$, the forms of $r_{k,1}$ and $r_{k,2}$ are different. Hence, to continuously simplify \textit{SP2}, we decompose it into another two subproblems---SP2\_1 with optimization variable $p_{k,1}$ and \textit{SP2\_2} with $p_{k,2}$.
\begin{subequations} \label{problem:sp2_1}
\begin{align} 
   \textit{\textbf{SP2\_1}}:&~{\min_{p_{k,1}} ~\alpha(\sum_{k=1}^K \frac{p_{k,1}d_{k,1}}{B_k \log_2(1+\frac{p_{k,1}g_{k,1}}{B_k N_k})})}, \tag{\ref{problem:sp2_1}}\\
    \text{subject to},& ~(\ref{constra:p_range}),~ r_{k,1} \ge r_{k,1}^{min},\label{constra:min_rk1}
\end{align}
\end{subequations}
\begin{subequations}\label{problem:sp2_2}
\begin{align}
   \textit{\textbf{SP2\_2}}:&~{\min_{p_{k,2}} ~\alpha(\sum_{k=1}^K \frac{p_{k,2}d_{k,2}}{B_k\log_2(1+\frac{p_{k,2}g_{k,2}}{B_kN_k+p_{k,1}g_{k,1}})})}, \tag{\ref{problem:sp2_2}}\\
    \text{subject to}, &~(\ref{constra:p_range}),~r_{k,2} \ge r_{k,2}^{min},\label{constra:min_rk2}
\end{align}
\end{subequations}
where $r_{k,1}^{min} = \frac{d_{k,1}}{T-\frac{\eta c_{k,1}D_{k,1}}{f_{k,1}}}$, and $r_{k,2}^{min}=\frac{d_{k,2}}{T-\frac{\eta c_{k,2}D_{k,2}}{f_{k,2}}}$.

In fact, \textit{SP2\_1} and \textit{SP2\_2} are two minimization sum-of-ratios problems, which are NP-complete \cite{freund2001solving} and challenging to solve. Therefore, to further make them solvable, we transform \textit{SP2\_1} and \textit{SP2\_2} into the epigraph form. 
To be concise and due to the same property of \textit{SP2\_1} and \textit{SP2\_2}, we write a general form as follows. Besides, for \textit{SP2\_1}, $\hat{p}_k=p_{k,1}$; for \textit{SP2\_2}, $\hat{p}_k=p_{k,2}$.

\vspace{-5pt}
\begin{subequations} \label{problem:sp2_v1}
\begin{align}
    \textit{\textbf{SP2\_{epi}}}:&~\min_{\hat{p}_{k}, \Gamma_{k}} ~\alpha\sum_{k=1}^K \Gamma_{k},\tag{\ref{problem:sp2_v1}} \\
    \text{subject to}, ~
     & (\ref{constra:p_range}),~\text{Constraint}_1,\notag
    %  & \frac{p_{k,1}d_{k,1}}{B_k \log_2(1+\frac{p_{k,1}g_{k,1}}{B_k N_k})} \le \Gamma_{k}, \label{constra:epi_upp1}
\end{align}
\end{subequations}
where $\Gamma_{k}$ is the auxiliary variable and the constraint

\begin{align}\label{constra:C1}
% \hspace{-7pt}
    \text{Constraint}_1 \!&=\! \begin{cases}\!
   SP2\_1 \!: (\ref{constra:min_rk1}),~\frac{p_{k,1}d_{k,1}}{\hat{r}_k} \le \Gamma_{k},\\~~~~~~~~\hat{r}_k=B_k \log_2(1+\frac{p_{k,1}g_{k,1}}{B_k N_k}),\\
   SP2\_2 \!: (\ref{constra:min_rk2}),\!\frac{p_{k,2}d_{k,2}}{\hat{r}_k} \!\le\! \Gamma_{k},\\~~~~~~~~\hat{r}_k=B_k\log_2(1+\frac{p_{k,2}g_{k,2}}{B_kN_k+p_{k,1}g_{k,1}}).
    \end{cases}
\end{align}

% \begin{subequations} \label{problem:sp2_v2}
% \begin{align} 
%     \textit{SP2\_v2}:&~\min_{p_{k,2}, \Phi_{k}}~\alpha\sum_{k=1}^K \Phi_{k}, \tag{\ref{problem:sp2_v2}} \\
%     \text{subject to}, \notag\\
%     & (\ref{constra:p_range}),~(\ref{constra:min_rk2}),\notag\\
%     &\frac{p_{k,2}d_{k,2}}{B_k\log_2(1+\frac{p_{k,2}g_{k,2}}{B_kN_k+p_{k,1}g_{k,1}})}\le \Phi_{k},
% \end{align}
% \end{subequations}
% where $\Phi_{k}$ is an auxiliary variable.

However, the above form is still \mbox{non-convex}. In \textit{SP2\_1}, since the only variable is $p_{k,1}$, it is obvious that $p_{k,1}d_{k,1}$ is convex, and $B_k \log_2(1+\frac{p_{k,1}g_{k,1}}{B_k N_k})$ is concave. This means each term of the objective function has the characteristics---the numerator is convex and the denominator is concave. In addition, \textit{SP2\_2} has the same characteristics. Because of such characteristics, we provide the following lemma to transform the subproblems into subtractive-form problems, which are equivalent to the original subproblems.

\begin{lemma} \label{lemma1}
If ($\hat{p}_{k}^*, \Gamma_{k}^*$) is the solution of \mbox{\textit{SP2\_{epi}}}, the following problem has $\hat{p}_{k}^*$ as its solution if there exist $\nu_{k} = \nu_{k}^*,~\Gamma_{k} = \Gamma_{k}^*$, $k=1,\cdots,K$. 
\vspace{-3pt}
\begin{align}
   \textit{\textbf{SP2\_sub}}:&~ \min_{\hat{p}_{k}}~\sum_{k=1}^K \nu_{k}[\hat{p}_{k}\hat{d}_{k}-\Gamma_{k}\hat{r}_k],\\
    % & + \mu_k(r_{k,1}^{min}-B_k\log_2(1+\frac{p_{k,1}d_{k,1}}{B_kN_k}))
    \text{subject to}:
    & (\ref{constra:p_range}),~\text{Constraint}_2,\notag
\end{align}
where 
\begin{align}
    \text{Constraint}_2=\begin{cases}
    SP2\_1\!:\!  (\ref{constra:min_rk1}),\\
    SP2\_2\!:\! (\ref{constra:min_rk2}).\notag
    \end{cases}
\end{align}
For \textit{SP2\_1}, $\hat{p}_k\hat{d}_k=p_{k,1}d_{k,1}$; for \textit{SP2\_2}, $\hat{p}_k\hat{d}_k=p_{k,2}d_{k,2}$.

Additionally, with $\nu_{k} = \nu_{k}^*$, $\Gamma_{k} = \Gamma_{k}^*$ and $\hat{p}_{k} = \hat{p}_{k}^*$, the following equations are satisfied:
\begin{align}
    \nu_{k}^* = \frac{\alpha}{\hat{r}_k},~
    \Gamma_{k}^* = \frac{\hat{p}_{k}^*\hat{d}_{k}}{\hat{r}_k},~k=1,\cdots, K. \label{nu_gamma_equation}
\end{align}
\end{lemma}
\textit{\textbf{proof}}.
The proof is provided by Lemma 2.1 in \cite{jong2012efficient}.

In brief, Lemma \ref{lemma1} proves that \mbox{\textit{SP2\_{epi}}} and \mbox{\textit{SP2\_sub}} are equivalent and have the same optimal solution. Hence, both subproblems \textit{SP2\_1} and \textit{SP2\_2} can be transformed into the form of \mbox{\textit{SP2\_sub}}.

Thus, to solve \mbox{\textit{SP2\_{epi}}}, we can solve \mbox{\textit{SP2\_sub}} to obtain $\hat{p}_{k}$ with given $\nu_{k}$ and $\Gamma_{k}$. 
Next, with the obtained $\hat{p}_{k}$, we can calculate the new $\nu_{k}$ and $\Gamma_{k}$ through Eq. (\ref{nu_gamma_equation}).

To solve \mbox{\textit{SP2\_sub}}, applying KKT conditions, we can get
% the Lagrange function needs to be given:
% \begin{align}
%     L_2(p_{k,1}) &= \nu_{k}[p_{k,1}d_{k,1}-\Gamma_{k}B_k\log_2(1+\frac{p_{k,1}g_{k,1}}{B_k N_k})] \notag\\
%     &+ \mu_{k}[r_{k,1}^{min} - B_{k}\log_2(1+\frac{p_{k,1}g_{k,1}}{B_kN_k})],
% \end{align}
% where $\mu_k$ is the non-negative Lagrange multiplier.

% Since applying KKT conditions is similar to the process in \textit{SP1}, we skip the deduction here due to the limited space. 

\begin{align} \label{sol_pk_hat}
    \hat{p}_{k}^* \!=\!
     \begin{cases} 
    &\hspace{-10pt}\frac{(\nu_{k}\gamma + \mu_k)B_k}{\nu_k d_{k,1} \ln{2}} \!-\! \Lambda,~\mu_k=0 \\
    &\hspace{-10pt} (2^{\frac{\hat{r}_{k}^{min}}{B_k}} \!-\! 1)\Lambda,~\mu_k>0,
    \end{cases}
\end{align}

% \begin{align} \label{sol_pk1}
%     p_{k,1}^* \!=\!
%      \begin{cases} 
%     &\hspace{-10pt}\frac{(\nu_{k}\gamma + \mu_k)B_k}{\nu_k d_{k,1} \ln{2}} \!-\! \frac{B_kN_k}{g_{k,1}},~\mu_k=0 \\
%     &\hspace{-10pt} (2^{\frac{r_{k,1}^{min}}{B_k}} \!-\! 1)B_kN_k/g_{k,1},~\mu_k>0,
%     \end{cases}
% \end{align}
where $\mu_k = [2^{\frac{\hat{r}_{k}^{min}}{B_k}}\Lambda(\ln{2})\nu_k \hat{d}_{k}/B_k-\nu_k \Gamma_k]^+$ and $[x]^+$ means $\max(0, x)$. In addition,
\begin{align}\hspace{-5pt}
\begin{cases}\hspace{-2pt}
    SP2\_1: \hat{p}_k = p_{k,1},\hat{r}_{k}^{min} = {r}_{k,1}^{min}, \Lambda=\frac{B_kN_k}{g_{k,1}},\\\hspace{-3pt}
    SP2\_2: \hat{p}_k = p_{k,2},\hat{r}_{k}^{min} = {r}_{k,2}^{min}, \Lambda=\frac{B_kN_k+p_{k,1}g_{k,1}}{g_{k,2}}\!.
\end{cases}\hspace{-8pt}
\end{align}

We have finished converting \textit{SP1} and \textit{SP2}. The algorithm for optimizing \textit{SP2} is listed in Algorithm \ref{algo:optimization_subp2_1}. The original algorithm is given in \cite{jong2012efficient} and is a Newton-like method.

In Algorithm \ref{algo:optimization_subp2_1}, we define $\boldsymbol{\varphi}(\Gamma_{k},\nu_{k})= [\varphi_1^T(\Gamma_{k}), \varphi_2^T(\nu_{k})]^T$, where
\begin{align}
    \varphi_1(\!\Gamma_{k}\!) \!=\! [-\hat{p}_{k}\hat{d}_{k}+\Gamma_k\hat{r}_k]^T\!,~
    \varphi_2(\!\nu_{k}\!) \!=\! [-\alpha \!+\! \nu_{k}\hat{r}_{k}]^T\!,~k \!\in\! [1, K].
\end{align}
The Jacobian matrices of $\varphi_1(\Gamma_{k})$ and $\varphi_2(\nu_{k})$ are

\begin{align}
    \varphi_1^{\prime}(\Gamma_{k}) = diag(\hat{r}_{k}),~ \varphi_2^{\prime}(\nu_{k}) = diag(\hat{r}_{k}),~k \in [1,K],
\end{align}
where $diag()$ stands for a diagonal matrix.

% We then define $\boldsymbol{\rho}(\Phi_k, u_k) = [\rho_1(\Phi_k), \rho_2(u_k)]$, where
% \begin{align}
%     \rho_1(\Phi_k) &= [-p_{1,2}d_{1,2}+\Phi_1 r_{1,2},\cdots,-p_{K,2}d_{K,2}+\Phi_K r_{K,2}]^T, \\
%     \rho_2(u_k) &= [-\alpha+u_1 r_{1,2},\cdots,-\alpha+u_K r_{K,2}]^T.
% \end{align}
% Besides, the Jacobian matrices of $\rho_1(\Phi_k)$ and $\rho_2(u_k)$ are
% \begin{align}
%     \rho_1^{\prime}(\Phi_k) = diag(r_{k,2}), ~k=[1,\cdots,K],\\
%     \rho_2^{\prime}(u_k) = diag(r_{k,2}),~k=[1,\cdots,K].
% \end{align}

\begin{algorithm}
\label{algo:optimization_subp2_1}
\caption{Optimization of \textit{SP2\_1}/\textit{SP2\_2}}
Initialize $j = 0$, $\zeta \in (0,1)$, $\epsilon \in (0,1)$ and feasible $\hat{p}_{k}^0$. \\
\Repeat{$\boldsymbol{\varphi}(\Gamma_{k},\nu_{k})=\boldsymbol{0}$ or reaching the maximum iteration number $J$}{
Calculate $(\nu^{(j)}_{k}, \Gamma_{k}^{(j)})$ according to Eq. (\ref{nu_gamma_equation}).

Get $(\hat{p}_{k}^{(j+1)})$ by calculating Eq. (\ref{sol_pk_hat}).

Let $i$ be the smallest integer satisfying
\begin{align} \label{i_integer}
    &|\boldsymbol{\varphi}(\Gamma_{k}+\zeta^i\sigma_{k,1}^{(j)}, \nu_{k}+\zeta^i\sigma_{k,2}^{(j)}| \notag\\ 
    & \le (1-\epsilon\zeta^i)|\boldsymbol{\varphi}(\Gamma_{k}^{(j)},\nu_{k}^{(j)})|,
\end{align}
where
\begin{align}
   & \sigma_{k,1}^{(j)} = -[\varphi_1^{\prime}(\Gamma_{k}^{(j)})]^{-1}\varphi_1(\Gamma_{k}^{(j)}),\notag\\
  & \sigma_{k,2}^{(j)} = -[\varphi_2^{\prime}(\nu_{k}^{(j)})]^{-1}\varphi_2(\nu_{k}^{(j)}), 
\end{align}

Update
\begin{align}
\hspace{-10pt}(\Gamma_{k}^{(j+1)},\nu_{k}^{(j+1)}) \!=\!  (\Gamma_{k}^{(j)} \!+\! \zeta^i\sigma_{k}^{(j)},~\nu_{k} \!+\! \zeta^i\sigma_{k,2}^{(j)}).
\end{align}

$j \leftarrow j+1$.
}
\end{algorithm}

% \begin{algorithm}
% \label{algo:optimization_subp2_2}
% \caption{Optimization of SP2\_2}
% Initialize $j = 0$, $\zeta \in (0,1)$, $\epsilon \in (0,1)$ and feasible $(p_{k,1}^0, p_{k,2}^0)$. \\
% \Repeat{$\boldsymbol{\rho}(\Phi_k, u_k)=\boldsymbol{0}$ or reaching the maximum iteration number $\mathcal{J}$}{
% Calculate
% \begin{align}
%     &(\Phi_k^{(j)}, u_k^{(j)}) \notag\\
%     &= (\frac{p_{k,2}^{(j)}d_{k,2}}{B_k\log_2(1+\frac{p_{k,2}^{(j)}g_{k,2}}{B_kN_k+p_{k,1}g_{k,1}})}, \frac{\alpha}{B_k\log_2(1+\frac{p_{k,2}^{(j)}g_{k,2}}{B_kN_k+p_{k,1}g_{k,1}})}), \notag\\
%     &\text{for $k=1,\cdots,K$}.
% \end{align}

% Obtain $p_{k,2}^{(j+1)}$ through solving  \mbox{\textit{SP2\_sub2}}.

% Let $i$ be the smallest integer which satisfies
% \begin{align}
%     |\boldsymbol{\rho}(\phi_k^{(j)}+\zeta^i\delta_{k,1}^{(j)}, u_k^{(j)}+\zeta^i\delta_{k,2}^{(j))}| \notag\\ \le (1-\epsilon\zeta^i)|\rho(\Phi_k^{(j)}, u_k^{(j)})|,
% \end{align}
% where
% \begin{align}
%     \delta_{k,1}^{(j)} &= -[\rho_1^{\prime}(\Phi_k^{(j)})]^{-1}\rho_1(\Phi_k^{(j)}),\\
%     \delta_{k,2}^{(j)} &= -[\rho_2^{\prime}(\nu_k^{(j)})]^{-1}\rho_2(\nu_k^{(j)}),
% \end{align}

% Update
% \begin{align}
%     (\Phi_k^{(j+1)}, \nu_k^{(j+1)}) = (\phi_k^{(j)}+\zeta^i\delta_{k,1}^{(j)}, u_k^{(j)}+\zeta^i\delta_{k,2}^{(j)}).
% \end{align}

% $j \leftarrow j+1$.
% }
% \end{algorithm}

\vspace{-2pt}
\subsection{Resource Allocation Algorithm} \label{subsec:re_al}
Based on Eq. (\ref{sol_pk_hat}), a resource allocation algorithm is proposed, which is an iterative optimization algorithm, as shown in Algorithm \ref{algo:re_al}. 
It first initially assigns a feasible solution set within the range of $\boldsymbol{f}$, $\boldsymbol{p}$ and $\boldsymbol{s}$. 
Next, iteratively solving \textit{SP1} (problem (\ref{subproblem1_dual})) and \textit{SP2} (\textit{SP2\_1} \& \textit{SP2\_2}) to obtain $(\boldsymbol{f}, \boldsymbol{s})$ and $\boldsymbol{p}$ respectively until convergence.

%% 后期商量，是否需要加入分配用户给信道的内容，fk和pk下标需要统一
\begin{algorithm}
\label{algo:re_al}
\caption{Resource Allocation Algorithm}
Initialize $\boldsymbol{S}^{(0)} \leftarrow (\boldsymbol{f}^{(0)}, \boldsymbol{s}^{(0)}, \boldsymbol{p}^{(0)})$ of problem
(\ref{problem:origin}). Iteration number $i=1$, the maximum number of iterations $\mathcal{M}$.

\While{$|\boldsymbol{S}^{(i)}-\boldsymbol{S}^{(i-1)}|> \varepsilon$ and $i\le \mathcal{M}$ }
{Solve problem (\ref{subproblem1_dual}), which is the dual problem of \textit{SP1}, to obtain ($\boldsymbol{f}^{(i)}$, $\boldsymbol{s}^{(i)}$) through CVX given $\boldsymbol{p}^{(i-1)}$.

Given $(\boldsymbol{f}^{(i)}, \boldsymbol{s}^{(i)})$, call Algorithm \ref{algo:optimization_subp2_1} ($\hat{p}_k = p_{k,1}$) to solve \textit{SP2\_1} and get $p_{k,1}^{(i)}$.

Given $p_{k,1}^{(i)}$, call Algorithm \ref{algo:optimization_subp2_1} ($\hat{p}_k = p_{k,2}$) to solve \textit{SP2\_2} to obtain $p_{k,2}^{(i)}$.

$\boldsymbol{p}^{(i)} \leftarrow [p^{(i)}_{1,1},\cdots,p^{(i)}_{K,1}, p^{(i)}_{1,2},\cdots,p^{(i)}_{K,2}]$.

$\boldsymbol{S}^{(i)} \leftarrow (\boldsymbol{f}^{(i)},~ \boldsymbol{s}^{(i)},~\boldsymbol{p}^{(i)})$.

$i \leftarrow i+1$.
}
% \vspace{-10pt}
\end{algorithm}

\vspace{-2pt}
\subsection{Time Complexity and Convergence Analysis}
\textbf{Time complexity}. 
We use floating point operations (flops) to analyze the time complexity. One flop is any mathematical operation (e.g., addition/subtraction/multiplication/division). 
Since steps 2--8 are the bulk of Algorithm \ref{algo:re_al}, we mainly analyze this part. Because Algorithm \ref{algo:optimization_subp2_1} is called in steps 4 and 5, we turn the view to Algorithm \ref{algo:optimization_subp2_1} first.

In Algorithm \ref{algo:optimization_subp2_1}, step 3, 4 and 6 take $\mathcal{O}(K)$ flops. 
Step 5 takes $\mathcal{O}((i+1)K)$, where $i$ is the smallest integer satisfying the inequality (\ref{i_integer}).
Therefore, Algorithm \ref{algo:optimization_subp2_1}'s time complexity is $\mathcal{O}((i+4)K)$.
Besides, in Algorithm \ref{algo:re_al}, step 3 solves problem (\ref{subproblem1_dual}) through CVX. Due to the use of the interior-point algorithm in CVX, the worst time complexity is $\mathcal{O}(K^{4.5}\log{\frac{1}{\epsilon}})$ \cite{luo2010semidefinite}.
Note that Algorithm \ref{algo:re_al} solves \textit{SP2} by calling Algorithm \ref{algo:optimization_subp2_1} iteratively. Thus, the time complexity of Algorithm \ref{algo:re_al} is $\mathcal{O}(K^{4.5}\log{\frac{1}{\epsilon}} + 2(i+4)K)$.

\textbf{Convergence Analysis}.
Algorithm \ref{algo:optimization_subp2_1} is called in Algorithm \ref{algo:re_al}, so first, we discuss the convergence of Algorithm \ref{algo:optimization_subp2_1}. 
The convergence proof is provided by Theorem 3.2 in \cite{jong2012efficient}. Additionally, according to Theorem 3.2 in \cite{jong2012efficient}, Algorithm \ref{algo:optimization_subp2_1} converges with a linear rate at any starting point $(\boldsymbol{\nu}^0, \boldsymbol{\Gamma}^0)$ and a quadratic convergence rate of the solution's neighborhood.
Therefore, Algorithm \ref{algo:re_al}, which iteratively solves \textit{SP1} to get $(\boldsymbol{f}, \boldsymbol{s})$ and \textit{SP2\_1} to get $\boldsymbol{p}$, will converge eventually.

\begin{algorithm}
\label{algorithm:Greedy}
\caption{Benchmark Greedy Algorithm}
Initialize $s=s_1$, $C_{min}=\infty$, $k\in [1,K]$

$P=[p_0,...,p_{10}],p_i = p^{min}+0.1i(p^{max}-p^{min})$, 

$F=[f_0,...,f_{10}],f_i = f^{min}+0.1 i(f^{max}-f^{min})$

\For{$f_{k,1}$ in $F$~and~$f_{k,2}$ in $F$}{
% \For{}{
\For{$p_{k,1}$ in $P$~and~$p_{k,2}$ in $P$}{
% \For{}{

The energy of channel $k$: $E \leftarrow$ Eq. (\ref{equa:trans_e})+Eq. (\ref{equa:cmp_e}).

The time consumption of channel $k$: $T\leftarrow$ Eq. (\ref{equa:trans_t})+ Eq. (\ref{equa:cmp_t}).

% E, time T of channel k separately according to Eq. (\ref{equa:trans_e})+ Eq. (\ref{equa:cmp_e}) and Eq. (\ref{equa:trans_t})+ Eq. (\ref{equa:cmp_t})\

$C \leftarrow \alpha E + \beta T$.

\If{$C < C_{min}$}{
Update $ C_{min} \leftarrow C, E_{out}\leftarrow E, T_{out}\leftarrow T $
}
% }
% }
}}
\textbf{output :} $C_{min}, E_{out}, T_{out}$
\end{algorithm}

\vspace{-2pt}
\section{Experimental Results}
In this section, we evaluate experimental results: 1) the effectiveness of our proposed algorithm in the joint optimization of energy and time consumption. 2) How the global model accuracy varies with the weight parameter $\gamma$.

% \vspace{-10pt}
\subsection{Parameter Settings}
The overall system settings are provided in Table \ref{tab:sys_para}.
% the number of channels $K=25$, the number of devices (i.e, users) $N=50$, the number of CPU cycles per sample $c_n$ is randomly assigned between $[1, 3]\times 10^4$. The path loss model is $128.1+37.6\log(\texttt{d})$ (the unit of \texttt{d} is kilometer). The standard deviation of shadow fading is $8$ dB. The noise power spectral density $N_0$ is -174 dBm/Hz.

% \vspace{-10pt}
\begin{table}[!b]
\caption{System Parameter Setting}\label{tab:sys_para}
\begin{tabular}{lllll}
\cline{1-2}
\textbf{Parameter}                                             & \textbf{Value}                                                                              &  &  &  \\ \cline{1-2}
The path loss model                                   & $128.1+37.6\log(\texttt{d (km)})$                                                  &  &  &  \\
The standard deviation of shadow fading               & $8$ dB                                                                             &  &  &  \\
Noise power spectral density $N_k$                         & $-174$ dBm/Hz                                                                      &  &  &  \\
The number of users $N$                               & $50$                                                                               &  &  &  \\
The number of channels $K$                            & $25$                                                                               &  &  &  \\
Total Bandwidth ($B$)                                 & $20$ MHz                                                                           &  &  &  \\
CPU frequency $f^{max}$, $f^{min}$                    & $2$ GHz, $0$ GHz                                                                   &  &  &  \\
The number of CPU cycles $c_n$                        & \begin{tabular}[c]{@{}l@{}}Randomly assigned in\\  $[1,3]\times 10^4$\end{tabular} &  &  &  \\
Transmission power $p^{max}$, $p^{min}$               & $12$ dBm, $0$ dBm                                                                  &  &  &  \\
The number of local iterations $\eta$                  & $10$                                                                               &  &  &  \\
The data size uploaded $d_n$                         & $28.1$ kbits                                                                       &  &  &  \\
The number of samples $D_n$                           & $500$                                                                              &  &  &  \\
Effective switched capacitance $kappa$                & $10^{28}$                                                                          &  &  &  \\
Video frame resolutions ($s_0$, $s_1$,  $s_2$, $s_3$) & ($100, 160, 320, 640$) pixels                                                      &  &  &  \\ \cline{1-2}
\end{tabular}
\end{table}
% \vspace{-10pt}

% Additionally, we set the number of local iteration $\xi$ as 10. The data size $d_n$ to upload and the samples $D_n$ for each device are set as 28.1 kbits and 500 separately.
% The effective switched capacitance $\kappa$ is $10^{28}$.
% The maximum CPU frequency $f^{max}$ is 2 GHz and the minimum frequency $f^{min}$ is 0 for each device. 
% The maximum transmission power $p^{max}$ is 20 dBm and the minimum one $p^{min}$ is 0 dBm. The total bandwidth $B$ is 20 MHz. 
% For the video resolution, we set $s_1=160$ pixels, $s_2=320$ pixels, $s_3=640$ pixels and $s_{0}=100$ pixels.
Recall that the constant $\xi$ equals $\frac{1}{s_{0}^2}$, and the accuracy metric $\mathcal{A}(s_{1,1},s_{1,2},...,s_{k,i})$ should be $\sum_{k=1}^K\sum_{i=1}^2A_{k,i}(s_{k,i})$.

The optimization objective is $\alpha\mathcal{E}+\beta\mathcal{T}-\gamma\mathcal{A}$
from (\ref{problem:origin}). We ``normalize" the weight parameters such that $\alpha+\beta=1$ by dividing each of them by $\alpha+\beta$. The intuition for doing this is that $\mathcal{T}$ and $\mathcal{A}$ are considered as "costs" and $\mathcal{A}$ is the "gain", then $\alpha+\beta=1$ means the coefficient for the cost part is 1.

\vspace{-2pt}
\subsection{System Parameters}
We have three weight parameters for the optimization problem: $\alpha$, $\beta$ and $\gamma$, and $\alpha+\beta=1$. 
If more focus is on energy consumption, the parameter $\alpha$ should be greater than $\beta$. 
If the delay is the objective to mainly optimize, $\beta$ should be set as a larger value.
Besides, the value of $\gamma$ affects object detection in a similar way. 
To explore the influence of three parameters separately, we first implement experiments under different $(\alpha,\beta)$ with fixed $\gamma$ and then show results under different $\gamma$ with fixed $(\alpha,\beta)=(0.5,0.5)$.

We compare three pairs of weight parameters $(\alpha,\beta)=(0.9,0.1),(0.5,0.5)$ and $(0.1,0.9)$ with random allocation strategy and one greedy algorithm (provided in Algorithm \ref{algorithm:Greedy}). 
$(\alpha,\beta)=(0.9,0.1)$ is carried out when devices are low-battery to save energy. 
$(\alpha,\beta)=(0.5,0.5)$ represents the ordinary situation to equally consider time and energy. Besides, $(\alpha,\beta)=(0.1,0.9)$ stresses the time-sensitive scenario.
 
\begin{figure}[!t]
\setlength{\abovecaptionskip}{0.cm}
\setlength{\belowcaptionskip}{-0.cm}
\centering
% \vspace{-0.3cm}
% \hspace{-10pt}
\includegraphics[width=0.48\textwidth]{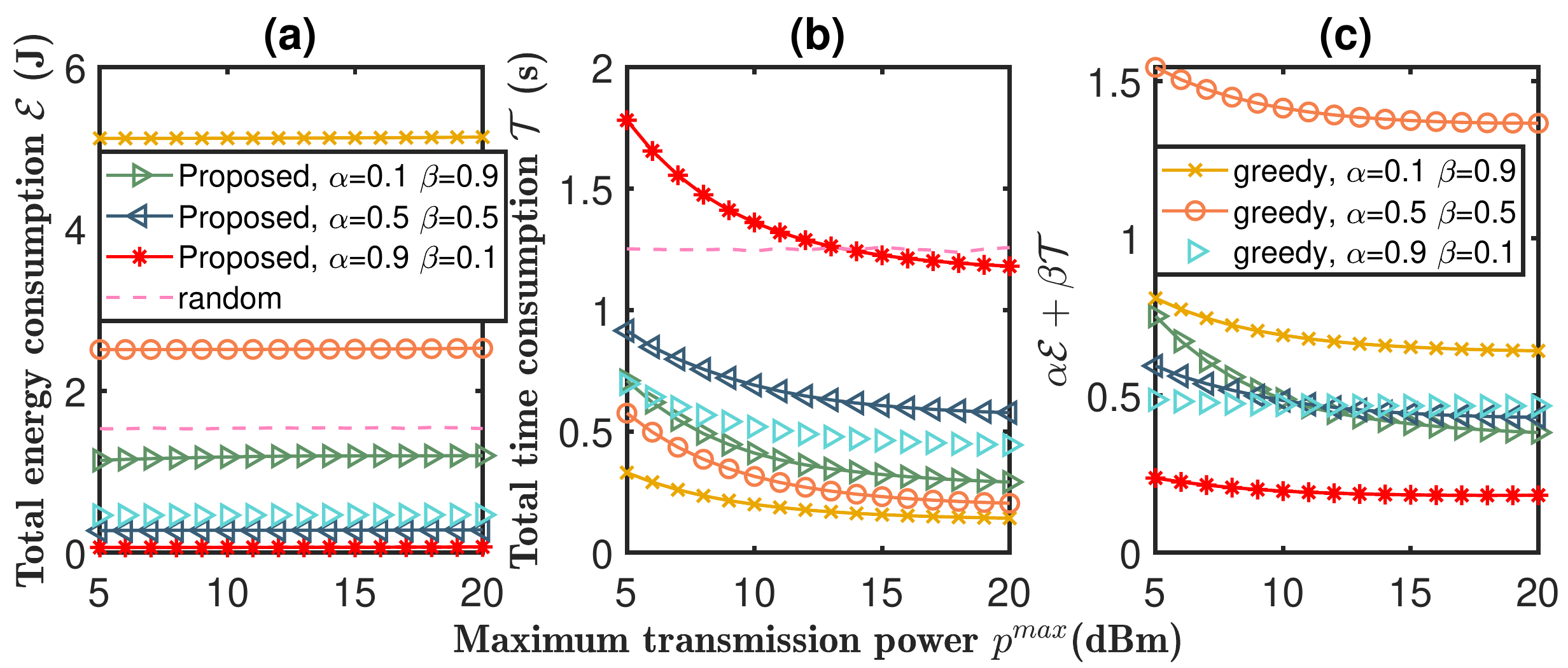}
\caption{Consumption under different maximum transmit power. $\gamma$=1. \vspace{-15pt}}
\label{fig:Consumption under different maximum transmit power}
\end{figure}

Fig. \ref{fig:Consumption under different maximum transmit power}(a), (b) and (c) show the total energy consumption $\mathcal{E}$, time consumption $\mathcal{T}$ and $\alpha \mathcal{E}+\beta \mathcal{T}$ under different maximum transmission power limits. It can be observed that as the maximum transmission power increases, $\mathcal{T}$ and $\alpha \mathcal{E} + \beta \mathcal{T}$ decrease while $\mathcal{E}$ slightly increases. This is because as the range of $p^{max}$ expands, there will be a more optimal solution to decrease the time consumption. 
Our proposed algorithm is superior to the random allocation strategy and greedy algorithm in terms of energy optimization and $\alpha \mathcal{E}+\beta \mathcal{T}$. In terms of total time consumption, the proposed algorithm performs worse than the greedy algorithm. When ($\alpha=0.1, \beta=0.9$), 
However, it is evident that the gap is much smaller than that of energy consumption, and this is why our proposed algorithm still performs better in terms of $\alpha \mathcal{E} + \beta \mathcal{T}$.

\begin{figure}[!t]
\setlength{\abovecaptionskip}{0.cm}
\setlength{\belowcaptionskip}{-0.cm}
\centering
\vspace{-0.3cm}
\includegraphics[width=0.46\textwidth]{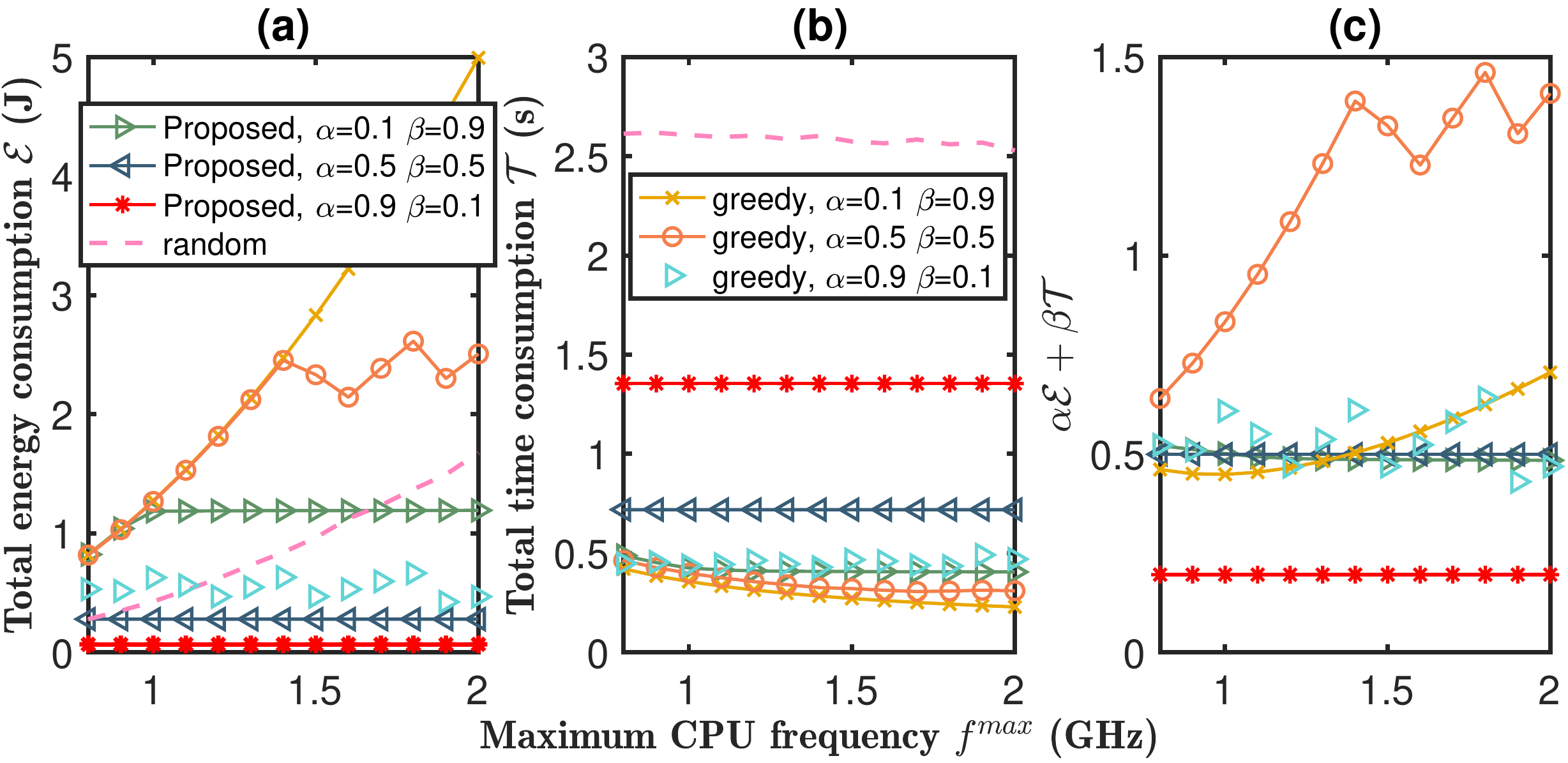}
\caption{Consumption under different maximum CPU frequency. $\gamma$=1.\vspace{-20pt}}
\label{fig:Consumption under different maximum CPU frequency}
\end{figure}

Moreover, Fig. \ref{fig:Consumption under different maximum CPU frequency} demonstrates the performance of different algorithms under different maximum CPU frequencies. From Fig. \ref{fig:Consumption under different maximum CPU frequency}(a), it can be seen the proposed algorithm performs much better than greedy algorithm in terms of energy consumption. Although the random strategy is slightly better than the green line (Proposed, $\alpha=0.1, \beta=0.9$) in terms of energy consumption, there is a huge gap in terms of time consumption as shown in Fig. \ref{fig:Consumption under different maximum CPU frequency}(b). Still, when maximum CPU frequency increases, the gap between proposed algorithm and greedy algorithm in terms of $\mathcal{E}$ becomes larger and larger. 
Although the performance of the proposed algorithm in terms of $\mathcal{T}$ is slightly worse than the greedy algorithm, the proposed algorithm is roughly more advantageous than the greedy algorithm in terms of $\alpha \mathcal{E} + \beta \mathcal{T}$.

% \vspace{-10pt}

\vspace{-2pt}
\subsection{Accuracy analysis}
\begin{figure}[!t]
\setlength{\abovecaptionskip}{0.cm}
\setlength{\belowcaptionskip}{-0.cm}
\centering
% \vspace{-0.1cm}
\includegraphics[width=0.88\linewidth]{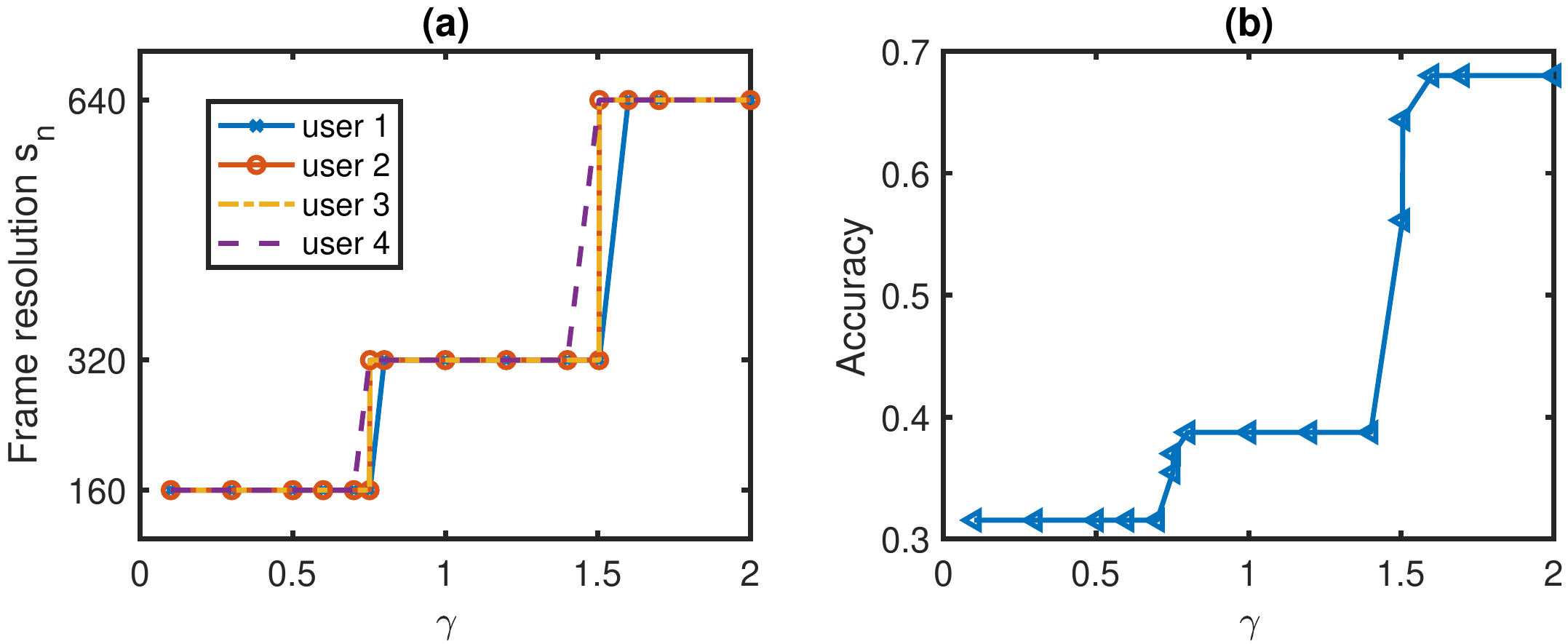}
\caption{Frame resolution and accuracy under different $\gamma$. Here $(\alpha,\beta)=(0.5,0.5)$. \vspace{-15pt}}
\label{fig:Resolution and Accuracy}
\end{figure}

To analyze the selection of the frame resolution for each device under different $\gamma$ and illustrate the result concisely, the number of users is set as 4. We fix the weight parameters $(\alpha,\beta)=(0.5,0.5)$ and choose different $\gamma$, to calculate corresponding optimal video resolution for each user as shown in Fig. \ref{fig:Resolution and Accuracy}(a) and illustrate the relationship between $\gamma$ and model accuracy in Fig. \ref{fig:Resolution and Accuracy}(b). YOLOv5m~\cite{glenn_jocher_2022_7002879}, which is one of the object detection architectures You Only Look Once (YOLO), is implemented for each user under federated learning setting. The dataset is COCO~\cite{lin2014microsoft}.
This experiment environment is on a workstation with three NVIDIA RTX 2080 Ti GPUs for computation acceleration.

There are 4 lines in Fig. \ref{fig:Resolution and Accuracy}(a) which represent the resolution choices for 4 users. Obviously, as $\gamma$ increases, which means the system will pay more attention on model accuracy, users are more inclined to choose higher video resolution. Because such selections of $s$ could bring prominent improvement for model accuracy, which could be noticed in Fig. \ref{fig:Resolution and Accuracy}(b). When $\gamma$ is lower than around 0.85, 4 users will always choose the lowest resolution $s_1=160$ and the model accuracy is only about 0.32. 
After $\gamma$ reach 0.85, users begin to pick resolution $s_2=320$ and the model accuracy increase to about 0.40. Besides, if $\gamma$ gets bigger, users will adopt the highest available resolution $s_3=640$ and so there is a dramatically improvement in accuracy, reaching approximately 0.68.

\section{Conclusion}
In this work, an FL-assisted MAR system via NOMA is proposed. We have explored the problem of joint optimization of energy, time and accuracy to allocate appropriate transmission power, computational frequency and video frame resolution for each device in the system. Through adjusting three weight parameters in the optimization problem, our proposed algorithm can be adapted to various scenarios. Time complexity and convergence analysis are also provided for the proposed resource allocation algorithm. In experimental results, it can be observed that our proposed algorithm is particularly effective in optimizing energy consumption compared to random allocation strategy and a benchmark greedy algorithm. Our paper also provides new insights into how federated learning and MAR can be applied to the Metaverse.

\section*{Acknowledgement}

This research is partly supported by the Singapore Ministry of Education Academic Research Fund under Grant Tier 1 RG90/22, RG97/20, Grant Tier 1 RG24/20 and Grant Tier 2 MOE2019-T2-1-176; and partly by the NTU-Wallenberg AI, Autonomous Systems and Software Program (WASP) Joint Project.

\ifCLASSOPTIONcaptionsoff
  \newpage
\fi

% \renewcommand{\baselinestretch}{1}
% \renewcommand*{\bibfont}{\small}
% \printbibliography
% \bibliographystyle{IEEEtran}
% \bibliography{references.bib}


\end{document}